 \documentclass[final,5p,times,twocolumn]{elsarticle}

\usepackage{amssymb}
\usepackage{amsmath}
\usepackage{braket}
\usepackage{bm}
\usepackage{hyperref}
\usepackage{orcidlink}

\journal{Physics Letters B}

\usepackage{xeCJK}
\biboptions{sort&compress}

\begin{document}

\begin{frontmatter}



\title{Prediction of deformed halo nuclei $^{43,45}$Si from multiple criteria based on structure and reaction analyses}


\author[AHNU]{C. Pan (潘琮)\orcidlink{0000-0003-3675-8238}}
\author[BUAA]{J. L. An (安嘉琳)}
\author[TUM,PKU]{P. Ring\orcidlink{0000-0001-7129-2942}}
\author[FZU]{X. H. Wu (吴鑫辉)\orcidlink{0000-0003-0237-5853}}
\author[RISP]{P. Papakonstantinou\orcidlink{0000-0002-7119-6667}}
\author[KNU]{M.-H. Mun\orcidlink{0000-0001-5439-3768}}
\author[CENS]{Y. Kim\orcidlink{0000-0003-0787-1485}}
\author[BUAA]{S. S. Zhang (张时声)\orcidlink{0000-0003-3926-7151}} \ead{zss76@buaa.edu.cn}
\author[INPC]{K. Y. Zhang (张开元)\orcidlink{0000-0002-8404-2528}} \ead{zhangky@pku.edu.cn}

\address[AHNU]{Department of Physics, Anhui Normal University, Wuhu 241000, China}
\address[BUAA]{School of Physics, Beihang University, Beijing 100191, China}
\address[TUM]{Physik-Department der Technischen Universit\"at M\"unchen, D-85748 Garching, Germany}
\address[PKU]{State Key Laboratory of Nuclear Physics and Technology, School of Physics, Peking University, Beijing 100871, China}
\address[FZU]{Department of Physics, Fuzhou University, Fuzhou 350108, Fujian, China}
\address[RISP]{Institute for Rare Isotope Science, Institute for Basic Science, Daejeon 34000, Republic of Korea}
\address[KNU]{Department of Physics, Kyungpook National University, Daegu 41566, Republic of Korea}
\address[CENS]{Center for Exotic Nuclear Studies, Institute for Basic Science, Daejeon 34126, Republic of Korea}
\address[INPC]{National Key Laboratory of Neutron Science and Technology, Institute of Nuclear Physics and Chemistry, China Academy of Engineering Physics, Mianyang, Sichuan 621900, China}

\begin{abstract}
Possible deformed neutron halos in silicon isotopes are investigated from both structure and reaction perspectives using the deformed relativistic Hartree-Bogoliubov theory in continuum (DRHBc) combined with the Glauber model.
The experimental neutron separation energies of silicon isotopes are well reproduced by the DRHBc theory. 
Multiple halo criteria are examined, including the global ones based on root-mean-square radii and density profiles, as well as the microscopic ones based on single-particle orbitals and their spatial distributions.
Calculations employing different density functionals and pairing strengths consistently indicate the emergence of $p$-wave neutron halos in $^{43,45}$Si, accompanied by pronounced shape decoupling between the halo and the core.
Moreover, the enhanced reaction cross sections and the narrow longitudinal momentum distributions of one-neutron removal residues provide additional evidence supporting the halo structures in $^{43,45}$Si.
\end{abstract}







\end{frontmatter}




\section{Introduction}

In 1985, a remarkable increase of the interaction cross section was observed in $^{11}$Li \cite{Tanihata1985PRL}, which was identified as a manifestation of the nuclear halo \cite{Hansen19887EPL,Kobayashi1988PRL}. 
In addition to halos, more exotic phenomena such as changes of nuclear magic numbers \cite{Ozawa2000PRL} and soft or pygmy resonances \cite{Nakayama2000PRL,Adrich2005PRL} have been discovered near drip lines. 
These exotic phenomena have presented notable challenges to traditional nuclear models, and have been drawing wide attention from both experimental and theoretical nuclear physicists over recent decades \cite{Ring1996PPNP,Jensen2004RMP,Tanihata2013PPNP,Otsuka2020RMP}.

The existence of halo structures has been experimentally identified or proposed in about 20 nuclides, as shown in Fig.~1 of Ref.~\cite{Zhang2023PRC-Na}. 
Halos are not rare phenomena but exist in a considerable number of the nuclei near the drip lines \cite{Tanihata2013PPNP}. 
Currently, the heaviest known halo nucleus is $^{37}$Mg \cite{Kobayashi2014PRL,Takechi2014PRC}. 
On the theoretical side, the description of the halos in light nuclei is already well under control \cite{Meng1996PRL,Poschl1997PRL,Rotival2009PRC1}. 
However, for medium and heavy nuclei, since the impact of one or two halo neutrons would be less prominent, existing definitions and tools are often too qualitative and the associated observables are incomplete \cite{Meng2015JPG}.
Over the past decades, characterizing the halo phenomena in medium-mass and heavier nuclei has been an active area of study in nuclear physics \cite{Meng1998PRL,Meng2002PRC,Rotival2009PRC1,Rotival2009PRC2,Meng2015JPG,Zhang2019PRC,Macchiavelli2022EPJA,Li2024PRC,Singh2024PLB,Papakonstantinou2025PRC,Singh2025arXiv}, and halos have been predicted in $^{39}$Na \cite{Zhang2023PRC-Na}, $^{42,44}$Mg \cite{Zhou2010PRC,Li2012PRC}, $^{40,42}$Al \cite{Zhang2023PRC-TRHBc,Zhang2025PRC-TRHBc}, \textit{etc}.

Silicon isotopes, ranging from the light region to the medium-mass region, exhibit a host of intriguing properties. 
The shell evolutions for proton at $Z=14$ and for neutron at $N=20,28$ in silicon isotopes have been drawing substantial experimental attention \cite{Fridmann2005Nature,Bastin2007PRL,Stroberg2014PRC,Stroberg2015PRC,Gade2019PRL,Gade2024PRC}, and the first proton density bubble was discovered in $^{34}$Si \cite{Mutschler2016NP}. 
These observations have also sparked considerable interest among theoretical nuclear physicists \cite{Meng1996PRL,Nowacki2009PRC,Utsuno2012PRC,Caurier2014PRC,Duguet2017PRC}. 
On the proton-rich side, the drip line of the silicon isotopic chain has been experimentally revealed as $^{22}$Si \cite{Xing2025PRL}. 
On the neutron-rich side, experimental efforts have pushed the boundary of observed Si isotopes to $^{46}$Si \cite{Notani2002PLB,Tarasov2007PRC,Yoshimoto2024PTEP}, which has not yet reached the drip line. 
Whether the discovered and undiscovered neutron-rich silicon isotopes exhibit halo structures remains unknown, due to the scarcity of experimental data. 

In this work, possible halos in silicon isotopes are explored with various halo criteria.
The microscopic structures are investigated based on the deformed relativistic Hartree-Bogoliubov theory in continuum (DRHBc) \cite{Zhou2010PRC,Li2012PRC}, and the reaction observables are evaluated using the Glauber model \cite{glauber1959lectures,Abu2003CPC}.
The DRHBc theory is briefly introduced in Section \ref{secth}. 
The numerical details are provided in Section \ref{secnum}. 
The results and discussion from both structure and reaction perspectives are presented in Section \ref{sechalo}. 
Finally, a summary is given in Section \ref{secsum}.

\section{The DRHBc theory} \label{secth}

In the theoretical description of nuclear halos, the effects of pairing correlation, continuum and deformation play significant roles \cite{Dobaczewski1984NPA,Meng1996PRL,Zhou2010PRC}. 
These effects could be properly considered by the state-of-the-art DRHBc theory in a self-consistent manner \cite{Zhou2010PRC,Li2012PRC}. 
The DRHBc theory has been applied in investigating the halos in boron \cite{Yang2021PRL,Sun2021PRC}, carbon \cite{Sun2018PLB,Sun2020NPA,Wang2024EPJA}, nitrogen \cite{Zhang2025PLB}, neon \cite{Zhong2022SCP,Pan2024PLB}, sodium \cite{Zhang2023PRC-Na}, magnesium \cite{Zhou2010PRC,Li2012PRC,Zhang2019PRC,Zhang2023PLB}, and aluminum \cite{Zhang2024PRC-Al,Papakonstantinou2025PRC} isotopes and many other exotic phenomena \cite{Sun2021SciB,Zhang2021PRC,Pan2021PRC,Xiao2023PLB,Zhao2023PLB,Mun2023PLB,Mun2024PRC,Mun2025PRC,Pan2025PRC,Qu2025NST,Zhang2025PRC}. 
Combining the DRHBc theory with the Glauber model, unified descriptions of the halo features in $^{37}$Mg and $^{15,19}$C are achieved from nuclear structure to reaction dynamics \cite{An2024PLB,Wang2024EPJA}. 
Recently, based on the DRHBc theory, a nuclear mass table including both deformation and continuum effects is under construction \cite{Zhang2020PRC,Zhang2022ADNDT,Pan2022PRC,Guo2024ADNDT}.
More relevant studies based on the DRHBc theory can be found in the recent review \cite{Zhang2025AAPPS}. 

For the details of the DRHBc theory, one can refer to Refs.~\cite{Li2012PRC,Zhang2020PRC,Pan2022PRC}, and here we outline its theoretical framework. 
The motion of nucleons is described by the relativistic Hartree-Bogoliubov (RHB) equation \cite{Kucharek1991ZPA}, 
\begin{equation} 
	\label{RHB}
	\begin{pmatrix} \hat{h}_D - \lambda_\tau & \hat{\Delta} \\ -\hat{\Delta}^* & -\hat{h}_D^* + \lambda_\tau \end{pmatrix} 
	\begin{pmatrix} U_k \\ V_k \end{pmatrix} = E_k \begin{pmatrix} U_k \\ V_k \end{pmatrix}, 
\end{equation}
where $\hat{h}_D$ is the Dirac Hamiltonian, $\hat{\Delta}$ is the pairing potential, $E_k$ is the quasiparticle energy, $U_k$ and $V_k$ are the quasiparticle wave functions, and $\lambda_\tau$ is the Fermi energy of neutron or proton $(\tau = n,p)$. 
The Dirac Hamiltonian $\hat{h}_D$, in coordinate space, reads
\begin{equation}
	h_D(\bm{r}) = \bm{\alpha}\cdot\bm{p} + V(\bm{r}) + \beta[M+S(\bm{r})],
\end{equation}
where $S(\bm{r})$ and $V(\bm{r})$ are scalar and vector potentials, respectively. 
The pairing potential $\hat{\Delta}$ reads
\begin{equation}
	\Delta(\bm{r}_1,\bm{r}_2) = V^{pp}(\bm{r}_1,\bm{r}_2) \kappa(\bm{r}_1,\bm{r}_2),
\end{equation}
where $V^{pp}$ is a density-dependent zero-range pairing force  \cite{Meng1998PRC2}, and $\kappa$ is the pairing tensor \cite{Ring1980NMBP}. 

The axial deformation and spatial reflection symmetry are assumed in the DRHBc theory, and the potentials and densities can thus be expanded in terms of the Legendre polynomials \cite{Price1987PRC,Xiang2023Symmetry},
\begin{equation}
	\label{elam}
	f(\bm{r}) = \sum_l f_l(r) P_l(\cos\theta), \qquad l = 0,2,4,\dots
\end{equation}
For the exotic nuclei close to drip lines, the continuum effect \cite{Dobaczewski1996PRC,Meng2006PPNP} is taken into account properly by solving the RHB equations \eqref{RHB} in a Dirac Woods-Saxon (DWS) basis \cite{Zhou2003PRC,Zhang2022PRC}. 
For a nucleus with odd number of neutron or proton, the blocking effect \cite{Ring1980NMBP} is considered via the equal-filling approximation \cite{Li2012CPL,Pan2022PRC}. 

\section{Numerical details} \label{secnum}

Numerical conditions of the present DRHBc calculations follow those employed in the construction of the global nuclear mass table \cite{Zhang2020PRC,Pan2022PRC}.
In coordinate space, a spherical box of radius $R_{\text{box}} = 20$~fm with a mesh size $\Delta r= 0.1$ fm is adopted.
The basis space is defined by an angular momentum cutoff of $J_{\max} = 23/2~\hbar$ and an energy cutoff of $E^+_{\text{cut}} = 300$ MeV for positive-energy states, with an equal number of negative-energy states included~\cite{Zhou2003PRC}. 
The Legendre expansion in Eq.~\eqref{elam} is truncated at $l_{\max} = 6$~\cite{Pan2019IJMPE}. 

Following the DRHBc mass table, the relativistic density functional PC-PK1 \cite{Zhao2010PRC}, a pairing strength $V_0 = -325 ~ \mathrm{MeV ~ fm^3}$, and a pairing window of 100 MeV are employed.
To examine possible density-functional dependence, additional calculations are performed using PC-L3R \cite{Liu2023PLB}, NL3$^*$ \cite{Lalazissis2009PLB}, and NL-SH \cite{Sharma1993PLB_NLSH}.
Meanwhile, to avoid ambiguities arising from the sensitivity to pairing parameters \cite{Karatzikos2010PLB}, three additional combinations of pairing strength and pairing window---each reproducing the odd-even mass differences for silicon isotopes---are also examined.
The details are provided in the Supplemental Material~\cite{Supp}.

 Recent studies have emphasized the significant role of three-body forces in determining root-mean-square (rms) radii, particularly in neutron-rich calcium isotopes \cite{Miyagi2020PRC,Huther2020PLB,Sommer2022PRL,Heinz2025PRC}. 
To examine the influence of three-body effects on our results, we performed additional calculations by explicitly adding an effective three-body force taken from the Skyrme interaction \cite{Vautherin1972PRC}, and then compared the corresponding results with those from the standard DRHBc calculations. 
It should be noted that such calculations are only an exploratory test, as the three-body force would inevitably be double-counted: the relativistic density functional already includes three-body effects in a phenomenological manner \cite{Anastasio1983PR}. 
The results show that the neutron halo is further enhanced by the three-body force.
Details of this test are provided in the Supplemental Material \cite{Supp}. 

\section{Results and discussion} \label{sechalo}

\begin{figure}[htbp]
	\centering
	\includegraphics[width=1.0\linewidth]{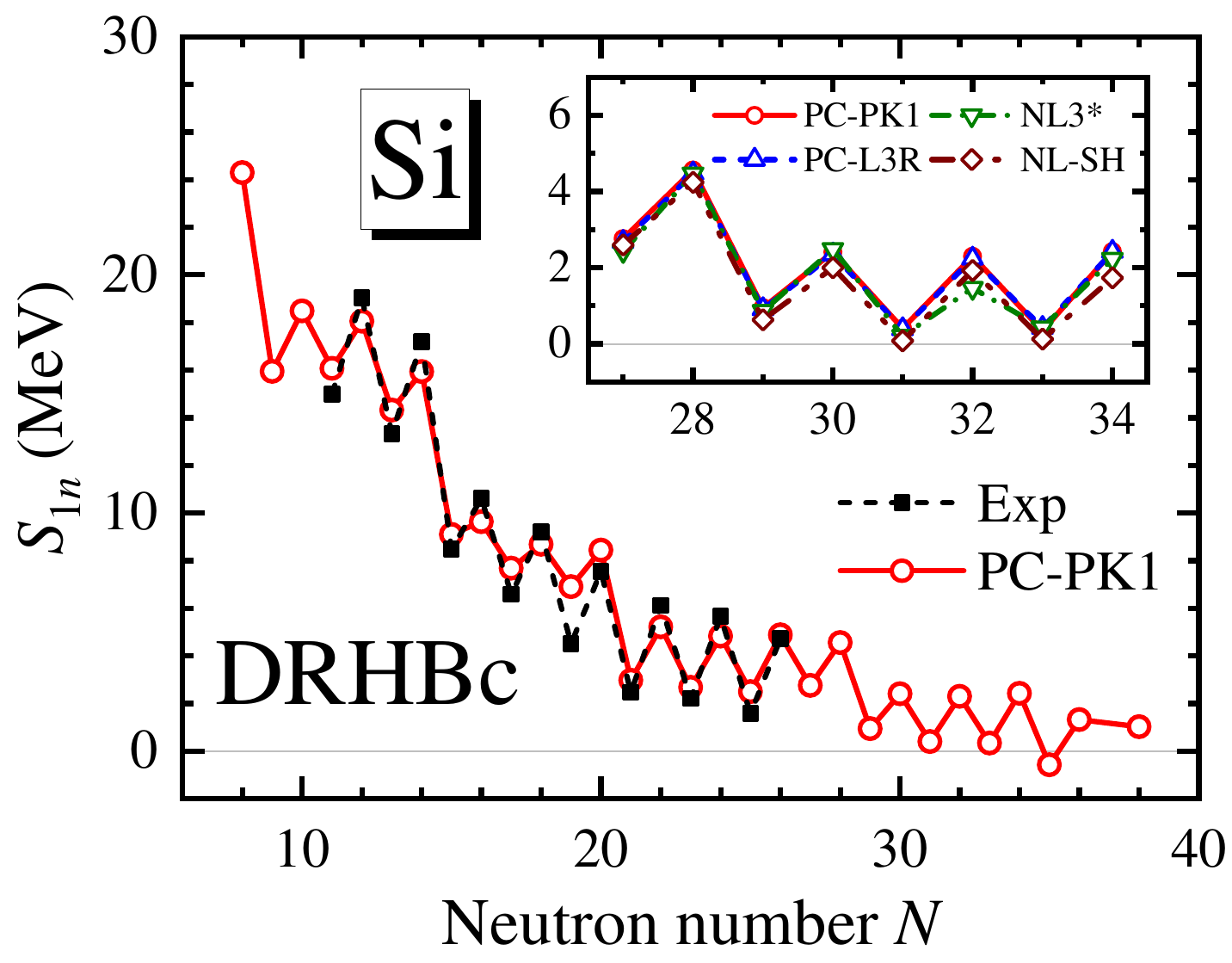}
	\caption{ One-neutron separation energy $S_{1n}$ as a function of neutron number $N$ in the DRHBc calculations with PC-PK1 for silicon isotopes, in comparison with the experimental data from AME2020 \cite{Wang2021CPC}. 
    The inset shows the $S_{1n}$ values obtained from DRHBc calculations with different density functionals, including PC-PK1, PC-L3R, NL3* and NL-SH, for silicon isotopes with $27 \leqslant N \leqslant 34$. }
	\label{f1}
\end{figure}

In Fig.~\ref{f1}, the one-neutron separation energy $S_{1n}$ for silicon isotopes from the DRHBc calculations with PC-PK1 is shown as a function of neutron number $N$, in comparison with the available data \cite{Wang2021CPC}. 
The data are reproduced with a rms deviation of $\sigma = 1.03$ MeV. 
Prominent staggering between even and odd $N$ is exhibited. 
The $S_{1n}$ value drops below 1 MeV for the odd-$N$ isotopes with $N\geqslant 29$, and becomes negative at $N=35$, marking the one-neutron drip line at $^{48}$Si. 
Notably, the small positive separation energies near the drip line may be signals of halo structures. 

The inset of Fig.~\ref{f1} compares the results for the neutron-rich isotopes $^{41\text{--}48}$Si obtained with the PC-L3R, NL3$^*$, and NL-SH density functionals to those calculated with PC-PK1.
All employed functionals yield mutually consistent results in both trend and magnitude.
In addition, neutron separation energies computed using different combinations of pairing strength and pairing window show only minor variations, particularly for the weakly bound silicon isotopes with $S_{1n}<1$~MeV \cite{Supp}.

\begin{figure}[htbp]
	\centering
	\includegraphics[width=0.8\linewidth]{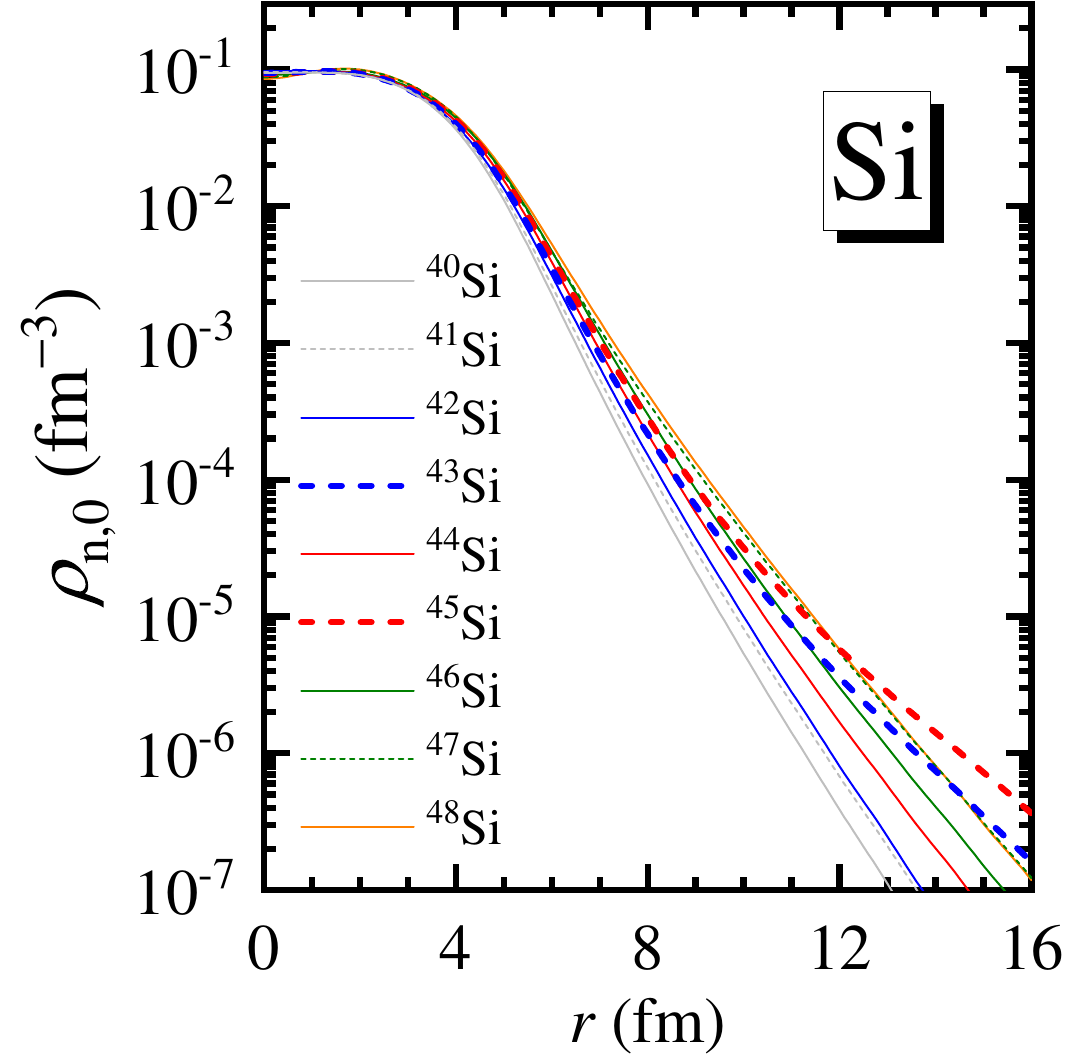}
	\caption{ Angle averaged neutron density distribution for $^{40-48}$Si in the DRHBc calculations with PC-PK1. }
	\label{f2}
\end{figure}

The presence of a halo is manifested by an extended tail of the density distribution. 
In Fig.~\ref{f2}, the angle averaged neutron density profiles for silicon isotopes near the one-neutron drip line from the DRHBc calculations with PC-PK1 are shown. 
At radii $r<8$ fm, the neutron density increases monotonically with $N$. 
In $^{43}$Si, the density becomes significantly larger than those of its even-even neighbors at $r>10$ fm, forming a diffused tail. 
A similar behavior is observed in $^{45}$Si, where the density exhibits an even more pronounced extension, exceeding those of all other isotopes at $r>12$ fm. 
Such large spatial extensions in density profiles, which are also obtained using other density functionals and pairing parameter sets for $^{43,45}$Si \cite{Supp}, provide strong indications of neutron halo formation.
In the following, we focus primarily on results calculated with the standard DRHBc mass-table numerical conditions.

To further explore possible halos in a more quantitative way, we analyze the halo criteria based on two global properties, which were proposed to characterize halos in not-so-light nuclei \cite{Rotival2009PRC1,Rotival2009PRC2,Zhang2023PRC-TRHBc,Zhang2025PRC-TRHBc}. 
These criteria do not require any \textit{a priori} division of the nucleus into a core and a halo. 

\begin{figure}[ht]
	\centering
	\includegraphics[width=0.8\linewidth]{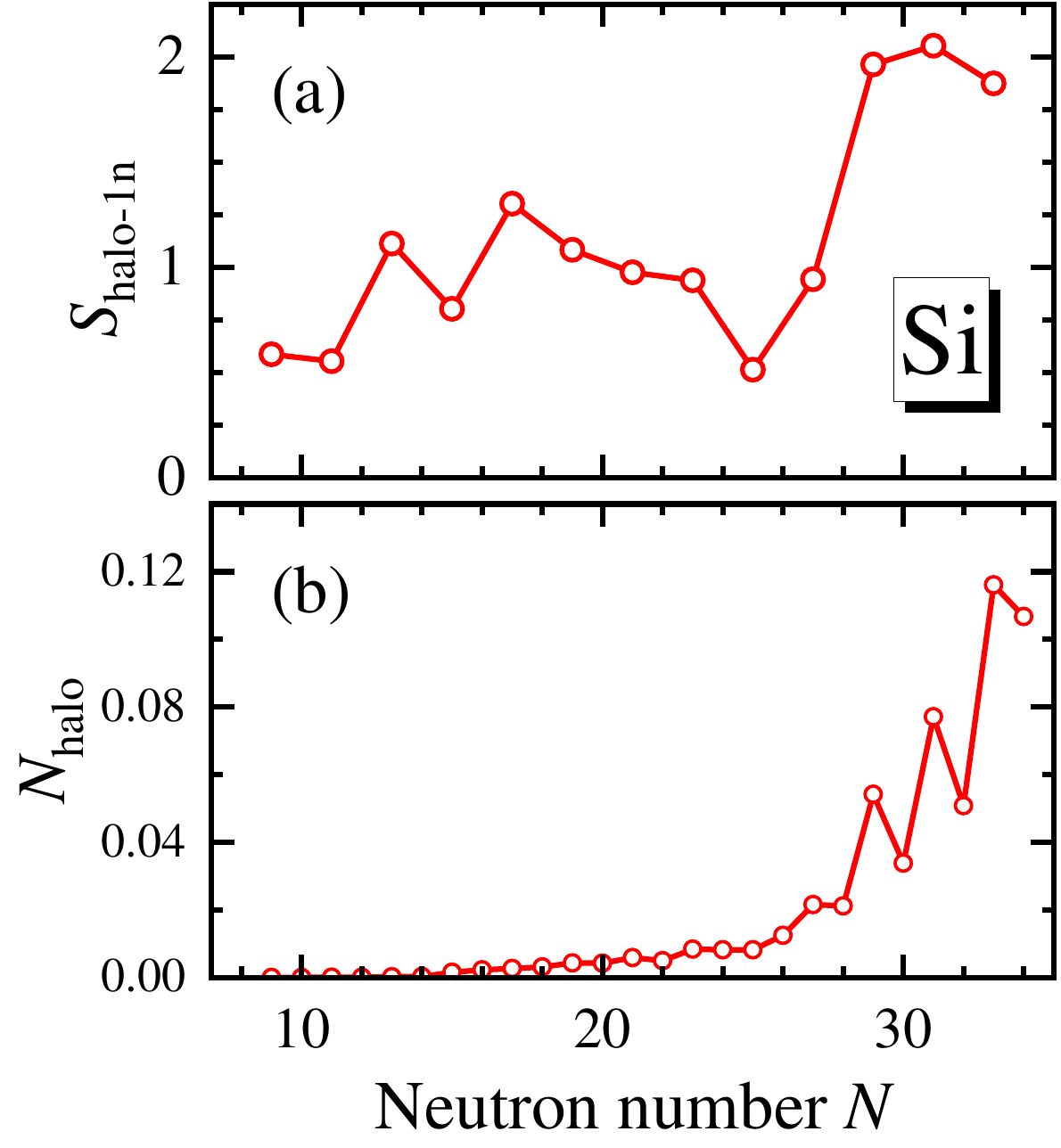}
	\caption{ (a) One-neutron halo scale $S_{\text{halo-1n}}$ and (b) halo parameter $N_{\text{halo}}$ for Si isotopes as functions of neutron number $N$ in the DRHBc calculations. See text for details. }
	\label{f3}
\end{figure}

\begin{itemize}
\item [(a)] \textit{Halo scale} ($S_{\text{halo}}$): 
This quantity was proposed in Ref.~\cite{Zhang2023PRC-TRHBc}, and an elevation of $S_{\text{halo}}$ might correspond to a halo nucleus \cite{Zhang2023PRC-TRHBc}. 
For the known halo nuclei and candidates, the halo scales extracted from experimental matter radii are generally larger than $2.0$ \cite{Zhang2025PRC-TRHBc}.
In this work, the one-neutron halo scale is calculated by \cite{Zhang2023PRC-TRHBc,Pan2025INPC}
\begin{equation}\label{Shalo}
	S_{\text{halo-1n}} = \frac{\Delta \tilde{R}_n}{\Delta R_n^{\text{emp}}} = \frac{\tilde{R}_n(N) - \tilde{R}_n(N-1)}{R_n^{\text{emp}}(N) - R_n^{\text{emp}}(N-1)}. 
\end{equation}
In Eq.~\eqref{Shalo}, $R_n^{\text{emp}} = r_0 N^{1/3}$ is the empirical radius, where $r_0=1.22$ fm is determined by the neutron radius of $^{34}$Si; 
$\tilde{R}_n = \left(1+5{\beta_2}^2 / {4\pi} \right)^{-1/2} R_n$ is the rms neutron radius after spherical reduction, which is proposed to avoid misleading results for the isotopes with sudden shape changes \cite{Pan2025INPC}, as the nuclear radius is appreciably impacted by deformation \cite{Pan2025PRC}. 
In Fig.~\ref{f3}(a), $S_{\text{halo-1n}}$ is near $1.0$ for $N\leqslant 27$, and rises abruptly to approximately $2.0$ at $^{43,45,47}$Si. 
Such enhancement of $S_{\text{halo-1n}}$, corresponding to a sudden increase of neutron radius (relative to the non-halo radius at the corresponding deformation), might be regarded as a signal of the halo phenomenon. 

\item [(b)] \textit{Average neutron number in the spatially decorrelated region} ($N_{\mathrm{halo}}$):
This quantity was proposed in Ref.~\cite{Rotival2009PRC1}, and a steep increase of $N_{\mathrm{halo}}$ may signal halo formation. 
It is defined as \cite{Rotival2009PRC1,Rotival2009PRC2}
\begin{equation}
	N_{\mathrm{halo}} = 4\pi \int_{r_0}^{\infty} \rho_{n,0}(r) r^2 \mathrm{d}r, 
\end{equation}
where $\rho_{n,0}(r)$ is the angle-averaged neutron density, and $r_0$ is proposed to determine the halo region \cite{Rotival2009PRC1}.
Here, to avoid numerical instability in high-order derivations, we take $r_0=r_{\max} + 2$~fm, where $r_{\max}$ corresponds to the maximum of $\partial^2 \lg\rho_{n,0} / \partial r^2$.
In Fig.~\ref{f3}(b), starting from the proton drip line, $N_{\mathrm{halo}}$ generally increases with $N$. 
Peaks of $N_{\text{halo}}$ occur at $^{43,45,47}$Si, consistent with their largest $S_{\text{halo-1n}}$ in Fig.~\ref{f3}(a). 

\end{itemize}

In the above discussion, both $S_{\text{halo-1n}}$ and $N_{\text{halo}}$ are enhanced in $^{43,45,47}$Si. 
For $^{43,45}$Si, the large $S_{\text{halo-1n}}$ and $N_{\mathrm{halo}}$ values are consistent with their spatial extensions in Fig.~\ref{f2}, and may correspond to neutron halos. 
For $^{47}$Si, the density distribution is somewhat less spatially extended compared to those of $^{43,45}$Si, yet it remains markedly more diffused than that of $^{46}$Si.
In addition, for $^{48}$Si, both $N_{\mathrm{halo}}$ and $\rho_{n,0}$ are close to the corresponding values for $^{47}$Si.
We note that $^{47}$Si and $^{48}$Si are spherical, and their potential halo structures will be examined in a separate study.

\begin{figure}[htbp]
	\centering
	\includegraphics[width=1\linewidth]{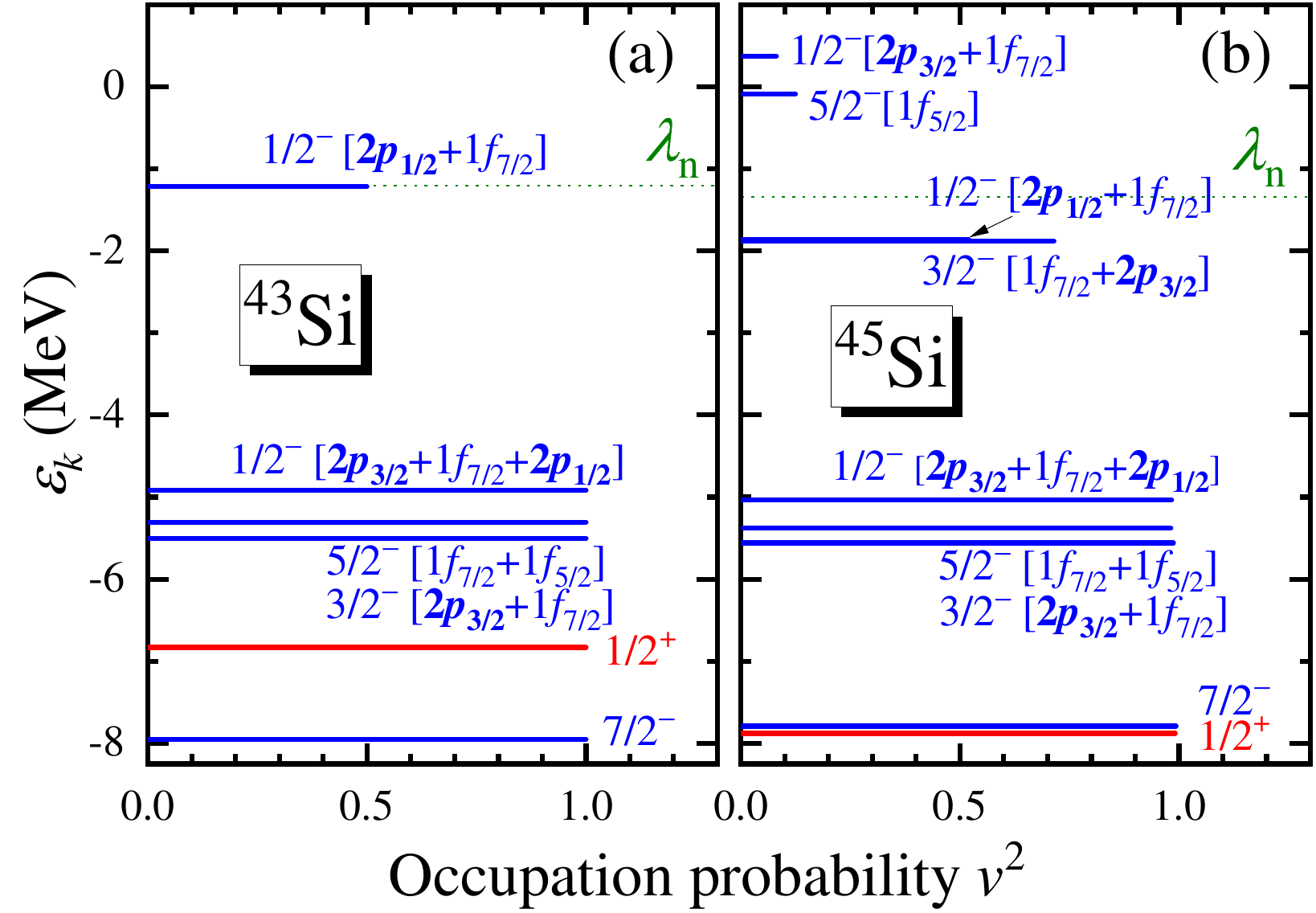}
	\caption{ Single-neutron levels around the continuum threshold in the canonical basis for $^{43}$Si (a) and $^{45}$Si (b) versus the occupation probability $v^2$ in the DRHBc calculations. 
	The levels are labeled by the quantum numbers $m^\pi$, and the main components in the DWS basis are given for the levels above $-6$ MeV. 
	The Fermi energy $\lambda_n$ is shown by a dotted line. 
	}
	\label{f4}
\end{figure}

It is generally considered that weakly bound low-$\ell$ orbitals and/or those embedded in the continuum are crucial to the formation of a nuclear halo \cite{Zhou2010PRC}. 
To examine the possible halos from this perspective, the single-neutron levels around the continuum threshold in the canonical basis for $^{43,45}$Si are shown in Fig.~\ref{f4}, with the main components in the DWS basis given for the levels above $-6$ MeV. 

For $^{43}$Si, as shown in Fig.~\ref{f4}(a), the valence neutron occupies a $1/2^-$ orbital, which is weakly bound at $\epsilon=-1.22$ MeV, with 58\% and 35\% contributions from $2p_{1/2}$ and $1f_{7/2}$, respectively.
A noticeable energy gap exists between this orbital and other more deeply bound ones with $\epsilon < -4$ MeV.
Consequently, the neutron density in $^{43}$Si can be naturally separated into two components: a ``core" formed by the deeply bound orbitals ($\epsilon < -4$ MeV) and the possible halo neutron.

For $^{45}$Si, as shown in Fig.~\ref{f4}(b), the valence neutron also occupies a weakly bound $1/2^-$ orbital, located at $\epsilon = -1.87$ MeV, with 62\% contribution from $2p_{1/2}$.
This orbital is nearly degenerate with a $3/2^-$ orbital at $\epsilon = -1.88$ MeV, with 28\% contribution from $2p_{3/2}$.
Owing to pairing correlations, several orbitals above the Fermi energy acquire nonzero occupation probabilities, including a $1/2^-$ orbital at $\epsilon = 0.37$ MeV with 50\% $2p_{3/2}$ components.
Similarly, a distinct energy gap emerges between the orbitals above $-2$ MeV---which contribute to the possible halo---and those below $-5$ MeV that form the core.

\begin{figure}[ht!]
	\centering
	\includegraphics[width=1\linewidth]{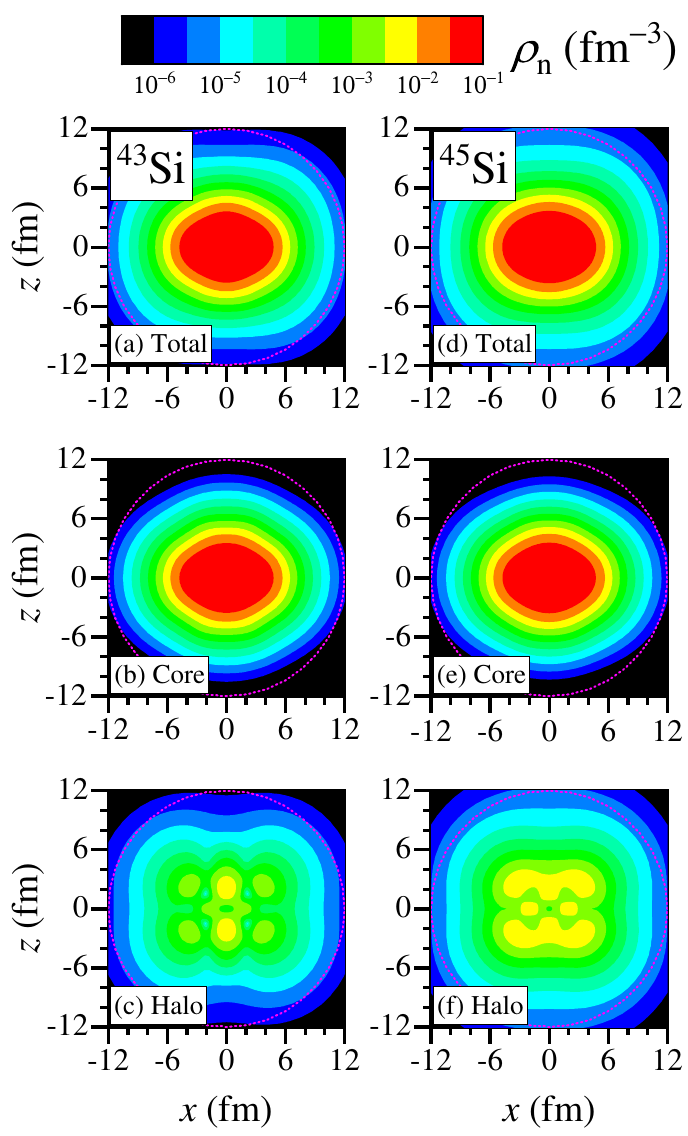}
	\caption{ Neutron density distributions with $z$-axis as the symmetry axis for $^{43}$Si [(a)-(c)] and $^{45}$Si [(d)-(f)]. Here, (a) and (d) show the total densities; (b) and (e) show the core densities; (c) and (f) show the halo densities.
	In each plot, a dotted circle is drawn to guide the eye.
	}
	\label{f5}
\end{figure}

To further characterize the spatial distribution and deformation of the possible neutron halos, Fig.~\ref{f5} displays the total neutron density distributions of $^{43,45}$Si, along with the individual contributions from the core and the possible halo.

For $^{43}$Si in Fig.~\ref{f5}(a), the neutron density displays an oblate shape with a quadrupole deformation parameter of $\beta_2 = -0.38$.
In Fig.~\ref{f5}(b), the core exhibits similar deformation ($\beta_{2,\text{core}} = -0.37$) and has a corresponding rms radius of $R_{n,\text{core}} = 3.83$ fm.
Figure~\ref{f5}(c) shows that the halo neutron is nearly spherical, with $\beta_{2,\text{halo}} = -0.01$.
The halo neutron presents a distinctly more extended density distribution, reflected also in its significantly larger rms radius $R_{n,\text{halo}} = 5.56$ fm compared to the core.
Taken together with the weak binding and dominant $p$-wave contribution, these results support the presence of a near-spherical neutron halo in $^{43}$Si, which is decoupled in shape from the oblate core.

For $^{45}$Si in Fig.~\ref{f5}(d), the neutron density is oblate, characterized by a quadrupole deformation parameter $\beta_2 = -0.31$.
In Fig.~\ref{f5}(e), the core exhibits similar oblate deformation ($\beta_{2,\text{core}} = -0.31$) and has an rms radius $R_{n,\text{core}} = 3.83$ fm.
In Fig.~\ref{f5}(f), the halo displays a near-spherical shape, with $\beta_{2,\text{halo}} = 0.01$, and possesses a more extended density distribution, corresponding to a larger radius $R_{n,\text{halo}} = 5.21$ fm.
These observations, together with the preceding analysis, indicate $^{45}$Si as also a deformed halo nucleus, in which the near-spherical halo is decoupled from the oblate core as well.

\begin{figure}[htbp]
	\centering
	\includegraphics[width=1.0\linewidth]{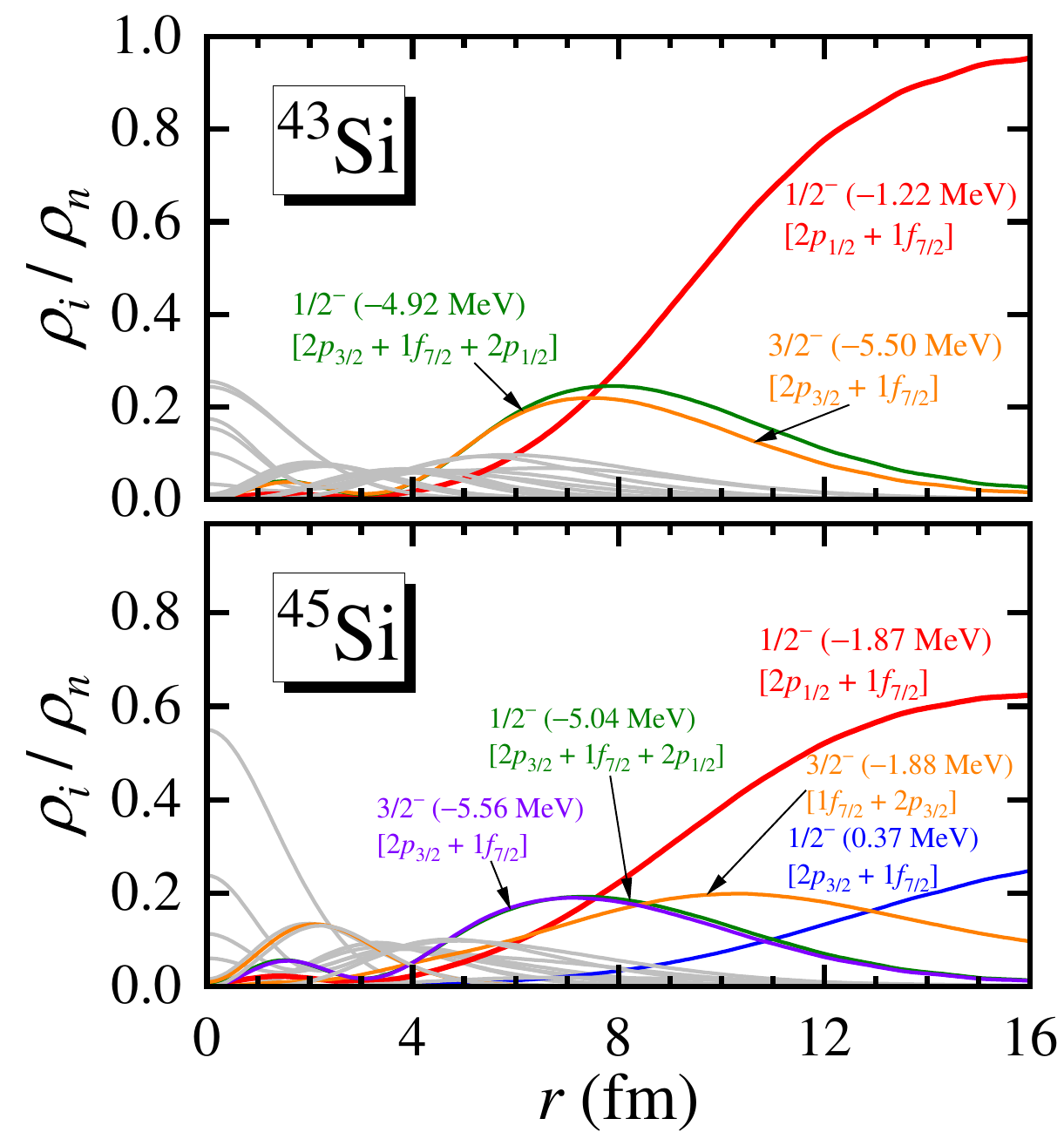}
	\caption{ Contribution of each single-neutron level in the canonical basis to the total neutron density (angle averaged) in $^{43}$Si (a) and $^{45}$Si as functions of the radius. 
	The levels with significant contributions ($\gtrsim 0.1$) in the asymptotic area are highlighted and their energies and main components are given. }
	\label{f6}
\end{figure}

To elucidate the halo formation mechanism in $^{43,45}$Si, in Fig.~\ref{f6}, the total neutron density is decomposed into contributions from individual single-neutron orbitals. 

For $^{43}$Si [Fig.~\ref{f6}(a)], the valence-neutron $1/2^-$ orbital dominates the density at large $r$.
This behavior arises from its predominant $2p_{1/2}$ component, which experiences a low centrifugal barrier and therefore extends far from the nuclear core.
In contrast, the $1/2^-$ and $3/2^-$ orbitals near $-5$~MeV, although also containing $p$-wave components, contribute only weakly to the density tail owing to their substantially larger binding energies.

For $^{45}$Si [Fig.~\ref{f6}(b)], similarly, the valence-neutron $1/2^-$ orbital that contains also a substantial $2p_{1/2}$ component provides the dominant contribution to the neutron density at large $r$.
Its nearly degenerate $3/2^-$ partner at $\epsilon = -1.88$ MeV contributes less in the outer region, owing to the smaller $p$-wave component.
Notably, the continuum $1/2^-$ orbital, despite its small occupation probability of $v^2 = 0.08$, exhibits an increasing contribution with $r$.
Its spatially extended density distribution arises from the combined effects of continuum coupling and strong $p$-wave character.
By contrast, the more deeply bound orbitals around $\epsilon \approx -5$~MeV contribute marginally at large $r$.

\begin{figure}[htbp]
	\centering
	\includegraphics[width=0.8\linewidth]{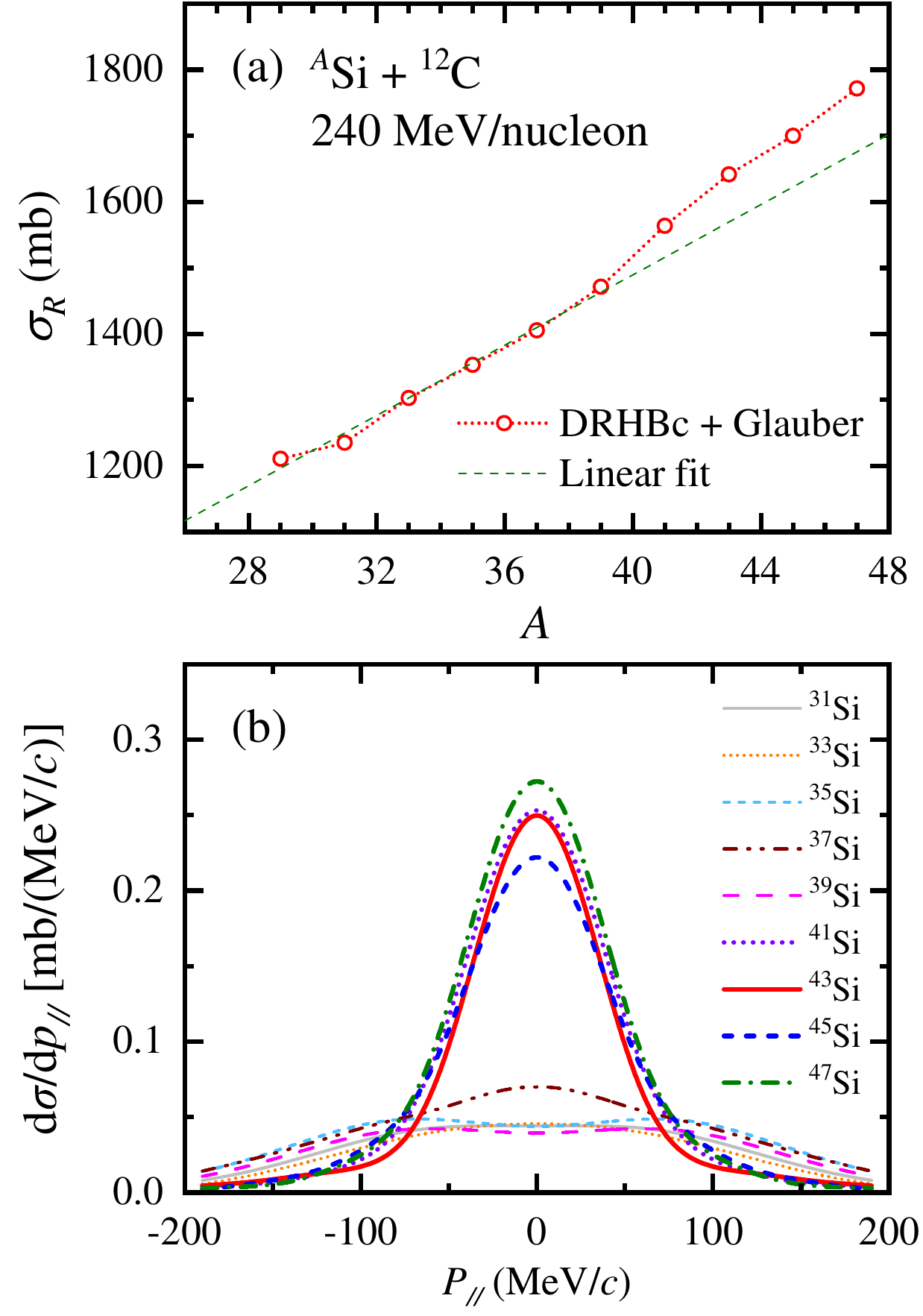}
	\caption{ (a) Reaction cross sections $\sigma_R$ of $^{29,31,...,47}$Si bombarding a $^{12}$C target at 240 MeV/nucleon evaluated using the Glauber model with inputs from the DRHBc model, in comparison with a linear fit using the results for $A\leqslant 39$. 
	(b) Inclusive longitudinal momentum distributions of residues after one-neutron removal reaction of $^{31,...,47}$Si bombarding a $^{12}$C target at 240 MeV/nucleon predicted with Glauber + DRHBc model. }
	\label{f7}
\end{figure}

Experimentally, evidence for halo structure includes a significant increase in the reaction cross section and a narrow momentum distribution of the residues \cite{Tanihata2013PPNP,Nakamura2020book}. 
Based on the structure inputs from the DRHBc theory, such reaction observables are evaluated using the Glauber model.
The theoretical framework of the Glauber model and validation for its reliability through the reproduction of experimental cross-section data for $^{28}$Si are provided in the Supplemental Material \cite{Supp}. 
It should be noted that the same parameters within the Glauber model are employed in the calculations for Si isotopes here and Mg isotopes in Ref.~\cite{An2024PLB}, where the reaction observables for the heaviest known halo nucleus $^{37}$Mg are reproduced rather well.
The results for one-neutron removal reactions of odd-mass silicon isotopes are illustrated in Fig.~\ref{f7}.

Figure \ref{f7}(a) presents the reaction cross sections $\sigma_R$ for $^{29,31,...,47}$Si bombarding a carbon target at 240 MeV/nucleon. 
For mass numbers $A=29\text{--}39$, $\sigma_R$ increases approximately linearly with $A$. 
A clear deviation from this linear trend emerges at $^{41}$Si, resulting in enhanced $\sigma_R$ values for $^{41,43,45,47}$Si relative to an extrapolation of the linear fit obtained for $A \leqslant 39$.
Figure \ref{f7}(b) displays the longitudinal momentum distributions for the residues following one-neutron removal from $^{31,33,...,47}$Si.
Notably, the distributions for $^{41,43,45,47}$Si exhibit narrow peak structures.
Quantitatively, their peak heights are approximately a factor of five larger than those for $A\leqslant 39$, while their full widths at half maximum are about $50~\mathrm{MeV}/c$, significantly smaller than the values ($\gtrsim 120~\mathrm{MeV}/c$) for $A\leqslant 39$.
Therefore, the neutron halos in $^{43,45}$Si predicted from the structure analyses are effectively supported by the reaction observables.

For $^{41}$Si, as shown in Figs.~\ref{f1} and \ref{f2}, the neutron separation energy exceeds 2 MeV, and the neutron density profile does not display an extended tail, instead lying between those of $^{40}$Si and $^{42}$Si.
Moreover, the valence-neutron orbital in $^{41}$Si does not exhibit a clear decoupling from the core, either in energy or in spatial distribution \cite{Supp}.
As a result, a simple ``core + $n$" picture is not well justified in $^{41}$Si, despite its implicit assumption in the Glauber-model calculation \cite{Abu2003CPC}, which yields a narrow momentum distribution due to the $p$-wave character of the valence neutron.
Consequently, the combined structural and reaction analyses do not support $^{41}$Si as a halo nucleus.
Further development of the Glauber model to incorporate full microscopic structure effects would therefore be desirable.

\section{Summary} \label{secsum}

In summary, possible deformed neutron halos in silicon isotopes have been investigated with the DRHBc theory.
The available experimental one-neutron separation energies of silicon isotopes are well reproduced. 
Multiple criteria for characterizing halos are examined, including global halo criteria---the halo scale and the average neutron number in the spatially decorrelated region---as well as the microscopic ones based on single-neutron orbitals and their density distributions.
The analyses, together with the weak binding of the valence neutron, indicate the emergence of $p$-wave neutron halos in $^{43,45}$Si, accompanied by pronounced shape decoupling from the oblate core. 
Reaction cross sections and longitudinal momentum distributions for silicon isotopes incident on a carbon target at 240 MeV/nucleon are further evaluated using the Glauber model with DRHBc structural inputs.
The enhanced cross sections and narrow momentum distributions provide additional support for halo formation in $^{43,45}$Si. 
Finally, the consistent results obtained with different density functionals---PC-PK1, PC-L3R, NL3$^*$, and NL-SH---and with different pairing parameter sets underscore the robustness of these predictions. 
As these extremely neutron-rich silicon isotopes are becoming accessible in experiments and more structural information is being revealed, $^{43,45}$Si are expected to set a new record as the heaviest halo nuclei in the near future.

\section*{Acknowledgements}

Helpful discussions with T. Nakamura, H. Sagawa, S. Q. Zhang, and members of the DRHBc Mass Table Collaboration are highly appreciated. %
This work was partly supported by the National Natural Science Foundation of China (Grant Nos. 12305125 and 12405134), the National Key Laboratory of Neutron Science and Technology (Grant No. NST202401016), Sichuan Science and Technology Program (Grant No. 2024NSFSC1356), and by the Deutsche Forschungsgemeinschaft (DFG, German Research Foundation) under Germany's Excellence Strategy EXC-2094-390783311, ORIGINS. 
The work of Y. K. was supported in part by the Institute for Basic Science (IBS-R031-D1). 
The work of M.-H. M. was supported by the National Research Foundation of Korea (NRF) grants funded by the Korean government (Ministry of Science and ICT) (Grant No. RS-2018-NR031074).



\renewcommand{\thefigure}{S\arabic{figure}}
\setcounter{section}{0}
\setcounter{figure}{0}

\newpage

\twocolumn[
\begin{center}
{\Large Supplemental Material for \\ Prediction of deformed halo nuclei $^{43,45}$Si from multiple criteria based on structure and reaction analyses}
\end{center}
]

\section{Dependence on pairing parameters } 

In the particle-particle channel, the density-dependent zero-range pairing force is adopted in the DRHBc calculations for silicon isotopes. 
The pairing parameters, \textit{i.e.}, the pairing cutoff (pairing window) $E_{\text{cut}}^{\text{qp}} = 100$ MeV and the pairing strength $V_0 = 325 ~ \mathrm{MeV ~ fm}^3$, are the same as those in the DRHBc mass table calculations, which were determined by reproducing the experimental odd-even mass differences for Ca and Pb isotopes \cite{supZhang2020PRC}. 

In Ref.~\cite{supKaratzikos2010PLB}, it is shown that beyond the region where the pairing parameters are adjusted, different $E_{\text{cut}}^{\text{qp}}$ values might produce inconsistent results. 
Therefore, here we examine whether changing pairing cutoff influences the halo phenomenon. 
Taking three cutoff values $E_{\text{cut}}^{\text{qp}}=100, 80$ and 60 MeV as examples, by minimizing the rms deviation of calculated odd-even mass differences from available data of neutron-rich Si isotopes, the sets $(E_{\text{cut}}^{\text{qp}}, V_0)=(100, 350), (80, 380)$ and $(60, 420)$ are obtained. 

\begin{figure}[htbp]
\centering
\includegraphics[width=.75\linewidth]{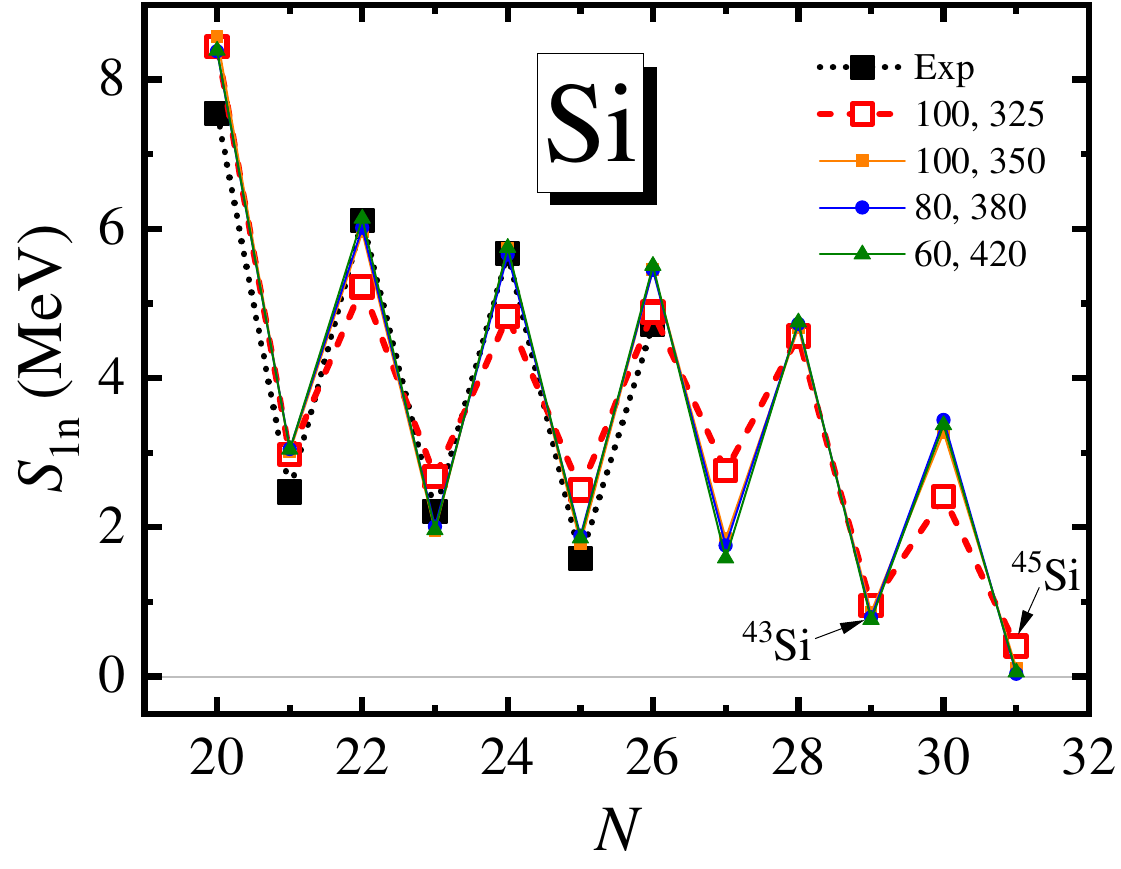}
\caption{One-neutron separation energy $S_{1n}$ as a function of neutron number $N$ for neutron-rich silicon isotopes, calculated by DRHBc with different pairing parameters $(E_{\text{cut}}^{\text{qp}}, V_0)$. 
The units of $(E_{\text{cut}}^{\text{qp}}, V_0)$ are $(\mathrm{MeV, MeV~fm}^3)$. 
The available experimental data from AME2020 \cite{supWang2021CPC} are shown for comparison. }
\label{FS1}
\end{figure}

Figure \ref{FS1} shows the one-neutron separation energies for neutron-rich silicon isotopes calculated using these pairing parameters, which are found essentially consistent with the $(100,325)$-results in the main text, particularly for $^{43,45}$Si.

\begin{figure}[htbp]
\centering
\includegraphics[width=\linewidth]{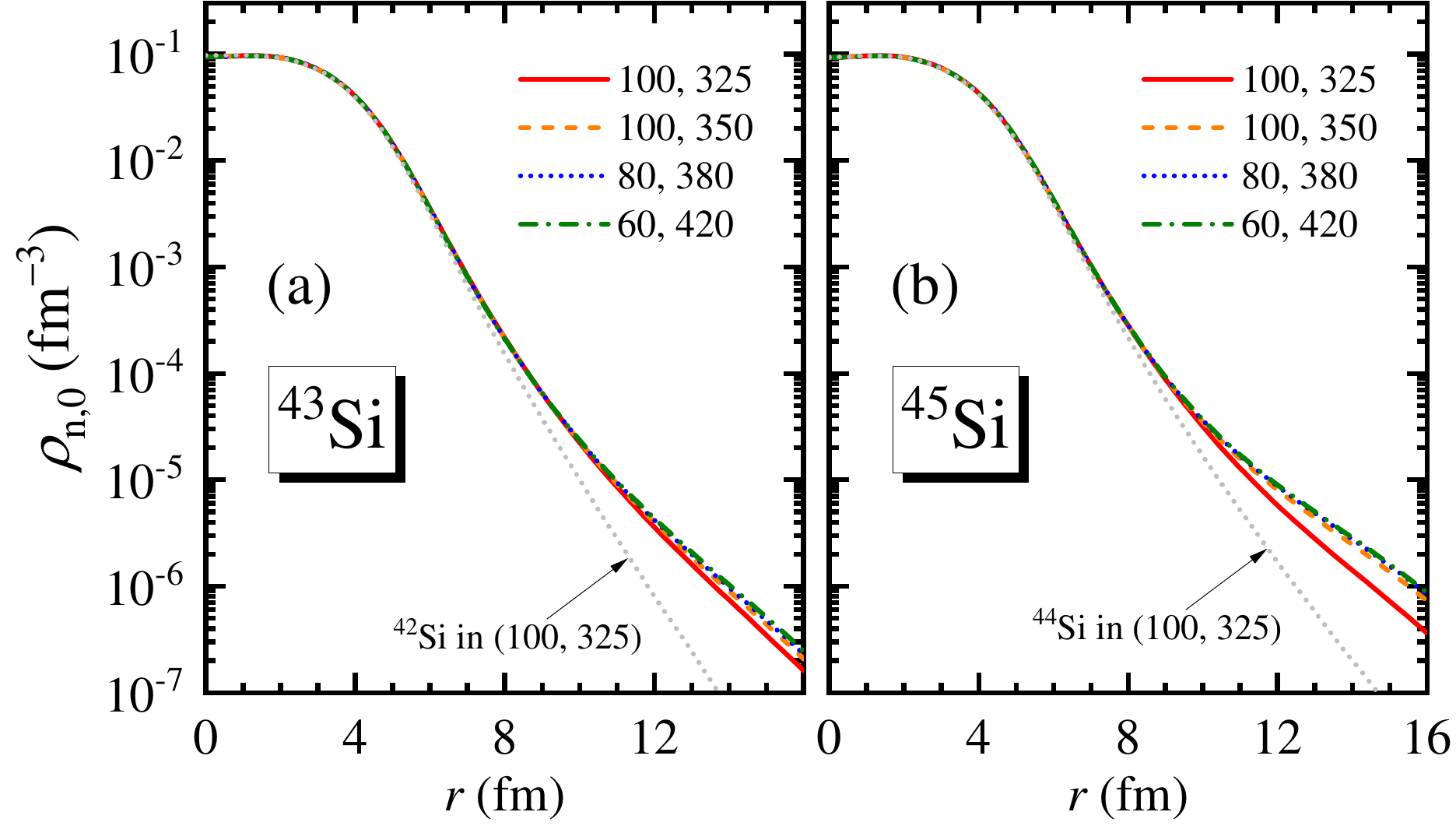}
\caption{Angle average neutron density distributions for $^{43}$Si (a) and $^{45}$Si (b), calculated by DRHBc with different pairing parameters $(E_{\text{cut}}^{\text{qp}}, V_0)$. The units of $(E_{\text{cut}}^{\text{qp}}, V_0)$ are $(\mathrm{MeV, MeV~fm}^3)$.  The densities of even-even neighbors are shown for comparison. }
\label{FS2}
\end{figure}

Figure \ref{FS2} shows the neutron density distributions for $^{43,45}$Si calculated using these new pairing parameters. 
The new results again coincide with the original ones---all are significantly larger than those of their even-even neighbors at large $r$. 
It is also noted that the results calculated with new pairing parameter sets exhibit slightly more extended distributions, which can be partially attributed to their smaller $S_{1n}$ values shown in Fig.~\ref{FS1}. 
Consequently, the prediction of neutron halos in $^{43,45}$Si is independent of the pairing cutoff. 

\section{Dependence on density functional }

In the particle-hole channel, four relativistic density functionals PC-PK1 \cite{supZhao2010PRC}, PC-L3R \cite{supLiu2023PLB}, NL3$^*$ \cite{supLalazissis2009PLB}, and NL-SH \cite{supSharma1993PLB_NLSH} are adopted in this work. 
The corresponding $S_{1n}$ values have been compared in Fig.~1 of the main text. 

\begin{figure}[htbp]
\centering
\includegraphics[width=\linewidth]{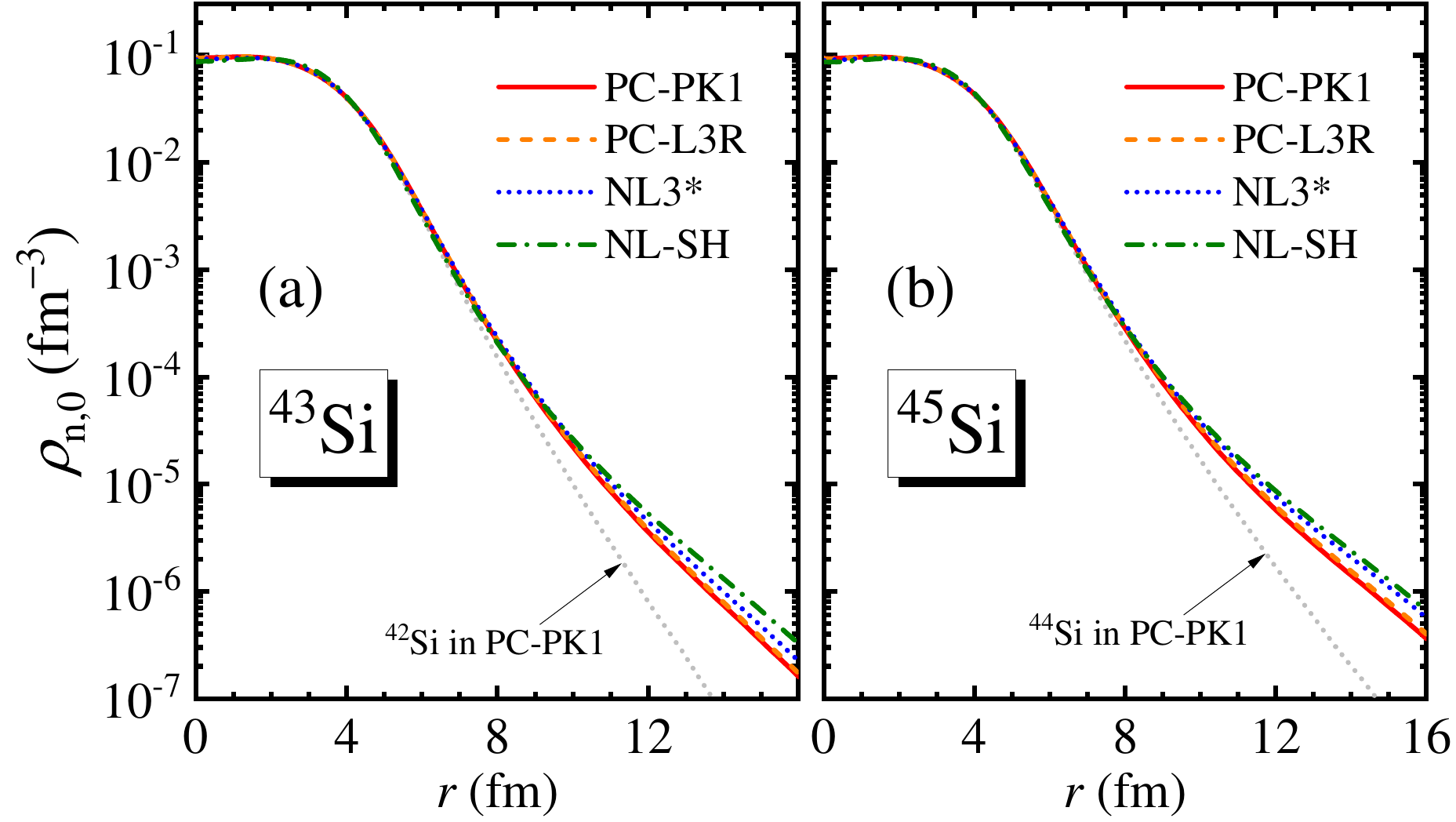}
\caption{Angle average neutron density distributions for $^{43}$Si (a) and $^{45}$Si (b), calculated by DRHBc with different density functionals. The densities of even-even neighbors are shown for comparison. }
\label{FS3}
\end{figure}

Figure \ref{FS3} shows the neutron density distributions of $^{43,45}$Si calculated with these density functionals. 
All the employed functionals yield results consistent with each other in both trend and magnitude. 
Therefore, the prediction of neutron halos in $^{43,45}$Si is independent of the choice of density functional. 

\section{ Discussion on $^{41}$Si }

In the main text, the reaction analysis shows halo-like behaviors for $^{41}$Si in Fig.~7, while it is not supported as a halo nucleus from the structure perspective.
This is because the Glauber-model calculation assumes a ``core~+~$n$'' structure, which is not well established for $^{41}$Si, as demonstrated below.

\begin{figure}[ht!]
\centering
\includegraphics[width=0.7\linewidth]{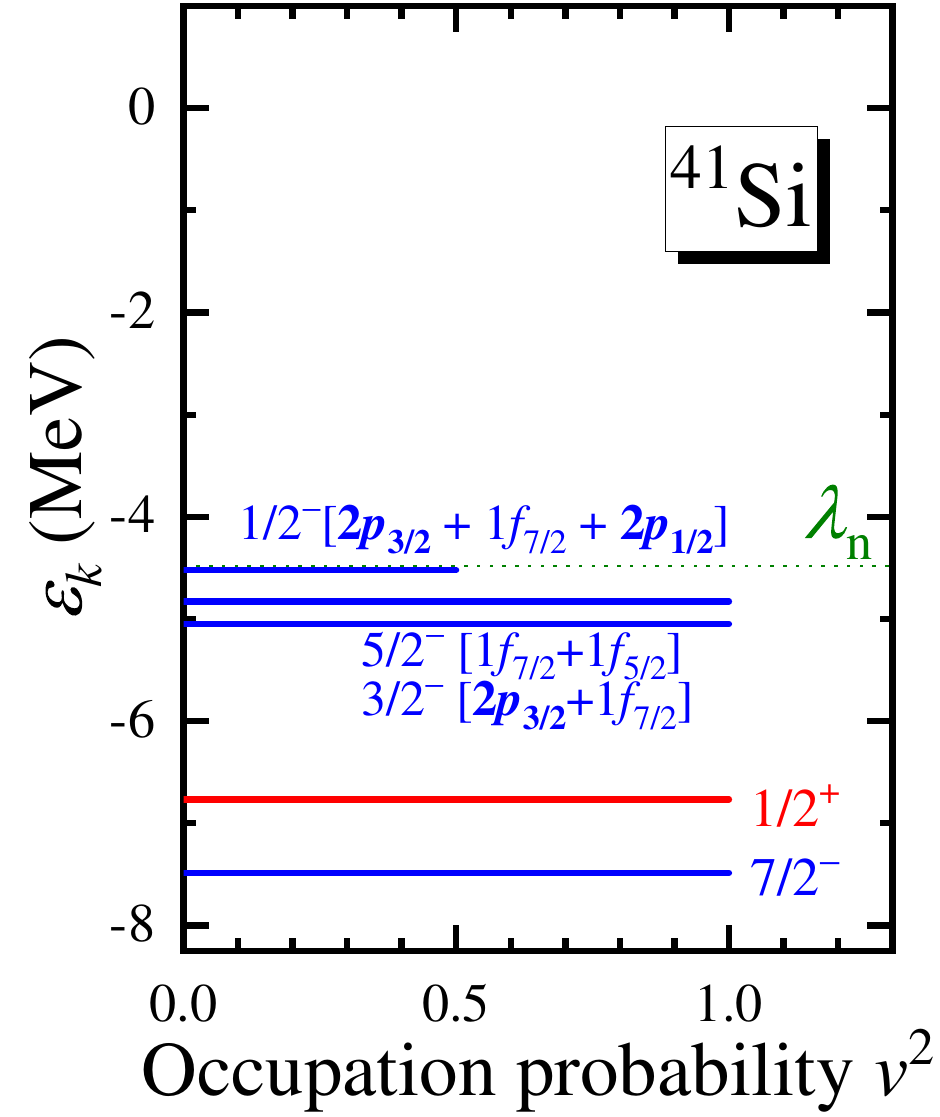}
\caption{ Single-neutron levels around the continuum threshold in the canonical basis for $^{41}$Si versus the occupation probability $v^2$ in the DRHBc calculations. The levels are labeled by the quantum numbers $m^\pi$, and the main components in the DWS basis are given for the levels above $-6$ MeV. The Fermi energy $\lambda_n$ is shown by a dotted line. }
\label{FS4}
\end{figure}

\begin{figure}[ht!]
\centering
\includegraphics[width=\linewidth]{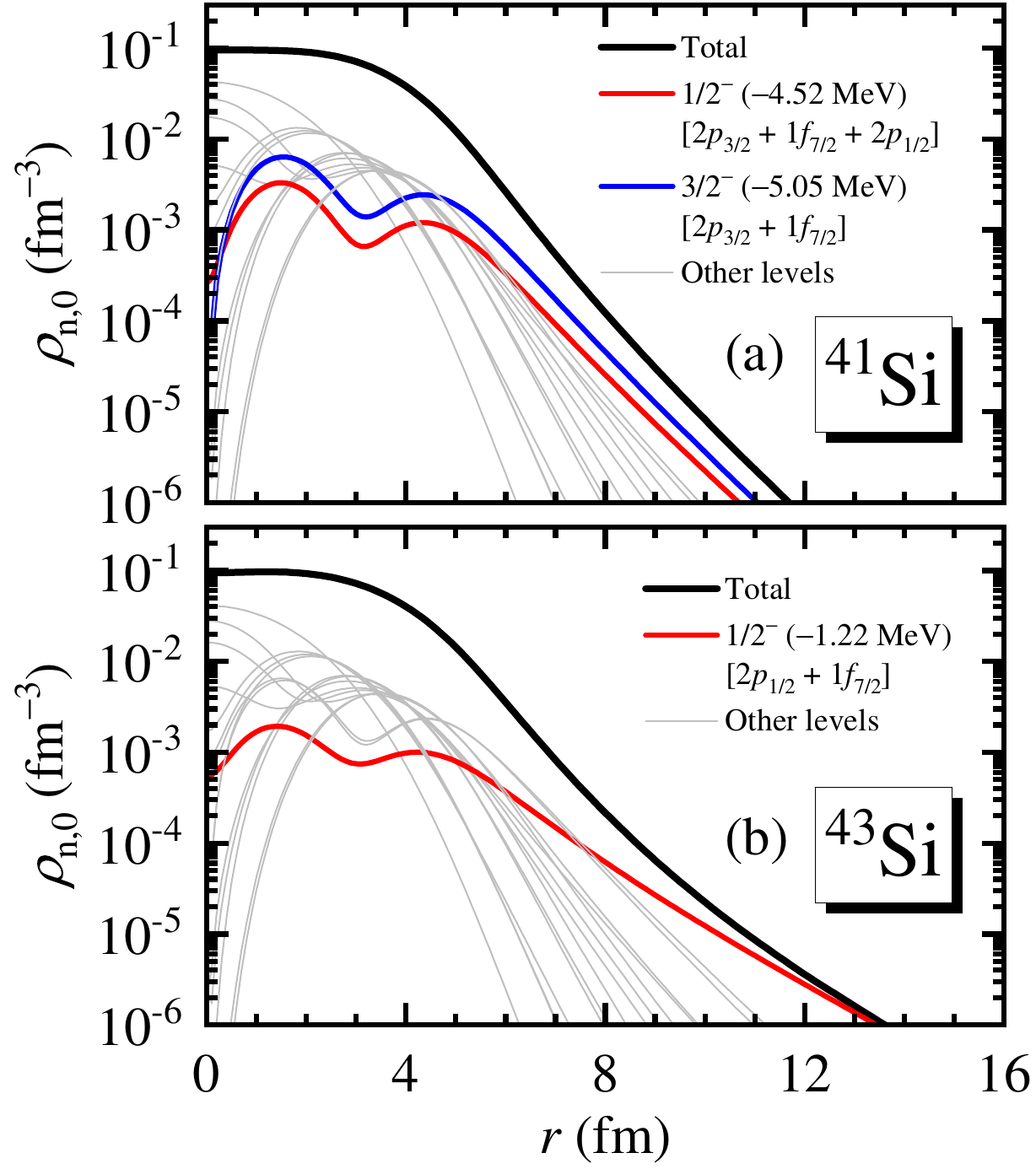}
\caption{ Angle-averaged densities of individual single-neutron orbitals in $^{41}$Si (a) and $^{43}$Si (b), along with the total neutron density, plotted as functions of the radial coordinate $r$. }
\label{FS5}
\end{figure}

Figure \ref{FS4} shows the single-neutron levels around the continuum threshold for $^{41}$Si. 
It can be found that the valence-neutron $1/2^-$ orbital of $^{41}$Si is not as loosely bound as those in $^{43,45}$Si (shown in Fig.~4 of the main text).
Meanwhile, it lies close to the $3/2^-$ orbital, with no sizable gap formed between the valence orbital and the core orbitals.
This is in sharp contrast to the cases of the predicted halo nuclei $^{43,45}$Si, where the halo orbital is well decoupled in energy from the core orbitals [Fig.~4 of the main text]. 

Figure \ref{FS5} shows the densities of individual single-neutron orbitals in $^{41}$Si, in comparison with those in $^{43}$Si.
For $^{41}$Si in Fig.~\ref{FS5}(a), the density of the valence-neutron orbital $1/2^-$ is smaller than that of the $3/2^-$ orbital.
Consequently, its contribution to the total neutron density is lower than that from the $3/2^-$ orbital.
This is clearly inconsistent with the classic picture of a halo nucleus, in which the valence nucleon(s) dominates the density in the asymptotic region.
In contrast, for $^{43}$Si in Fig.~\ref{FS5}(b), the valence-neutron orbital is well decoupled in spatial distribution from the core orbitals, forming an extended tail of the neutron density.
Therefore, from the structure perspective, $^{41}$Si is not suggested as a halo nucleus and does not exhibit a ``core~+~$n$'' structure.

\section{Influence of three-body forces}

Recent studies have emphasized the significant role of three-body forces in determining rms radii, particularly in neutron-rich systems such as calcium isotopes \cite{supHuther2020PLB,supMiyagi2020PRC,supSommer2022PRL,supHeinz2025PRC}. 
Therefore, it might be interesting to examine how such three-body effects may influence the present results within the DRHBc framework. 

In Brueckner-Hartree-Fock studies of nuclear matter, it was shown that relativistic density functionals do not require explicit terms of three-body forces \cite{supAnastasio1983PR}. 
It is a general assumption that three-body forces are already embedded in the functionals. 
Since practical functionals employ sophisticated density dependences, disentangling the influence of explicit three-body forces from the additional correlations encoded in the density dependence is extremely difficult and beyond the scope of the present work. 

However, to gain a more intuitive understanding of the effects of three-body force, we performed an exploratory study by comparing the results from the calculations with and without effective three-body contributions. 
We have adopted a three-body force that in Hartree-Fock calculations takes a density-dependent form \cite{supVautherin1972PRC}:
\begin{equation}
	v_{12} = \frac{1}{6} t_3 (1+P_\sigma) \delta(\bm{r}_1 - \bm{r}_2) \rho\left(\frac{\bm{r}_1+\bm{r}_2}{2}\right) . 
\end{equation}
By varying the strength parameter $t_3$ we can study the influence of three-body forces on halo behavior. 

We note that this is only an exploratory study: 
On the one hand, since the three-body effects have already been phenomenologically included in the relativistic density functional, simply adding $v_{12}$ would lead to double counting. 
On the other hand, to be as self-consistent as possible, we would need not only to add a proper three-body term, but also to globally rescale all remaining parameters in the relativistic density functional so as to achieve the same ground-state properties as before. 
Such an adjustment entails a huge computational cost and is beyond the scope of the present work. 
Therefore, we employ small values for $t_3$, aiming to explore the influence of three-body force on halo-related properties. 

\begin{figure}[htbp]
\centering
\includegraphics[width=\linewidth]{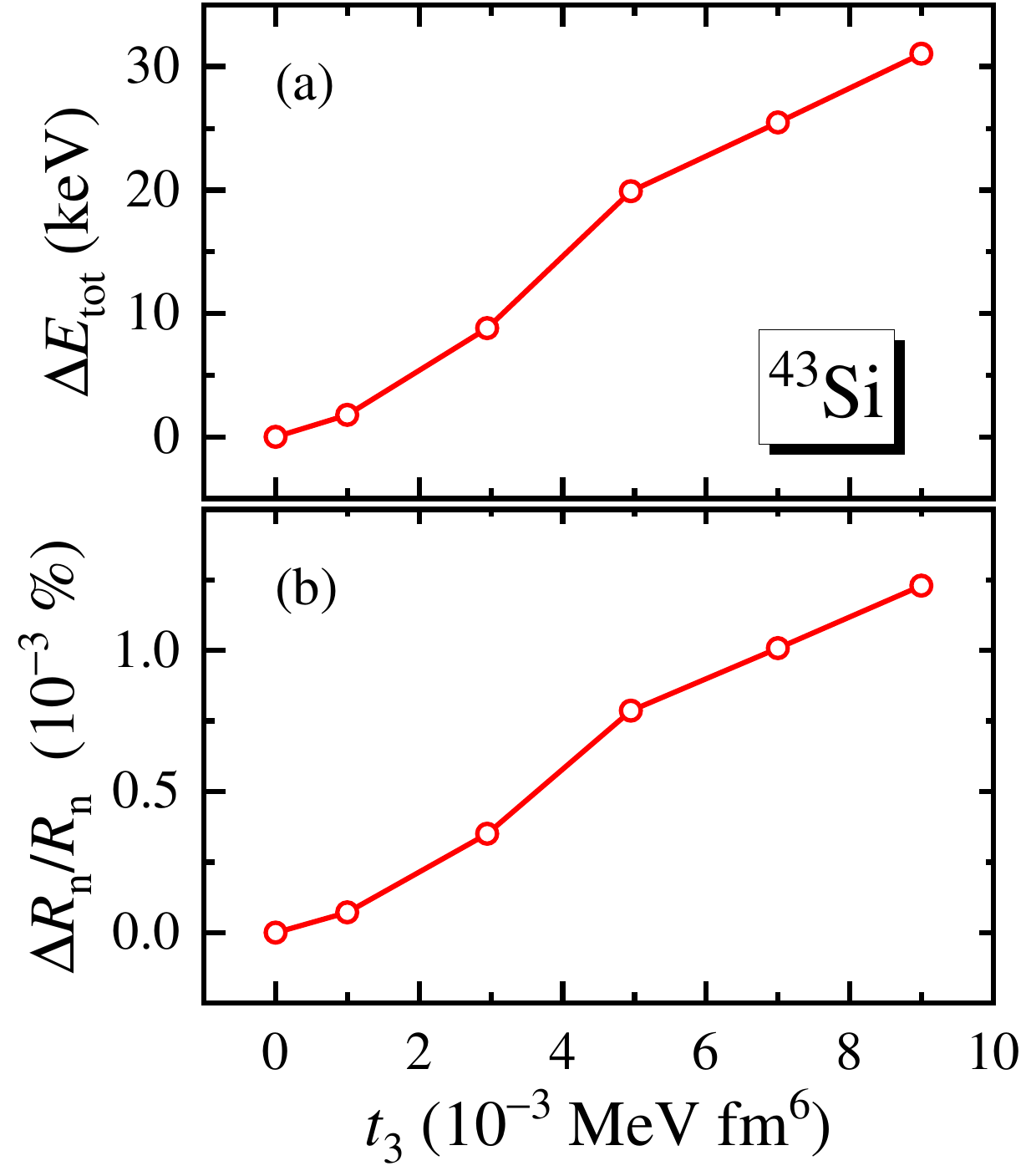}
\caption{(a) Difference of total energies $\Delta E_{\text{tot}}$ and (b) relative difference of neutron rms radii $\Delta R_n / R_n$ between the exploratory DRHBc calculations with and without phenomenological three-body force for $^{43}$Si. }
\label{TBF_diff_E}
\end{figure}

Taking $^{43}$Si as an example, we have performed ``DRHBc + three-body force'' calculations with different $t_3$ values. 
Figure \ref{TBF_diff_E} shows the influences of three-body force on the total energy and neutron rms radius. 
As $t_3$ increases, both $\Delta E_{\text{tot}}$ and $\Delta R_n / R_n$ increase monotonically. 
This indicates that three-body force makes the nucleus less bound and gives rise to a larger neutron rms radius, which favors the formation of a neutron halo. 

\begin{figure}[htbp]
\centering
\includegraphics[width=\linewidth]{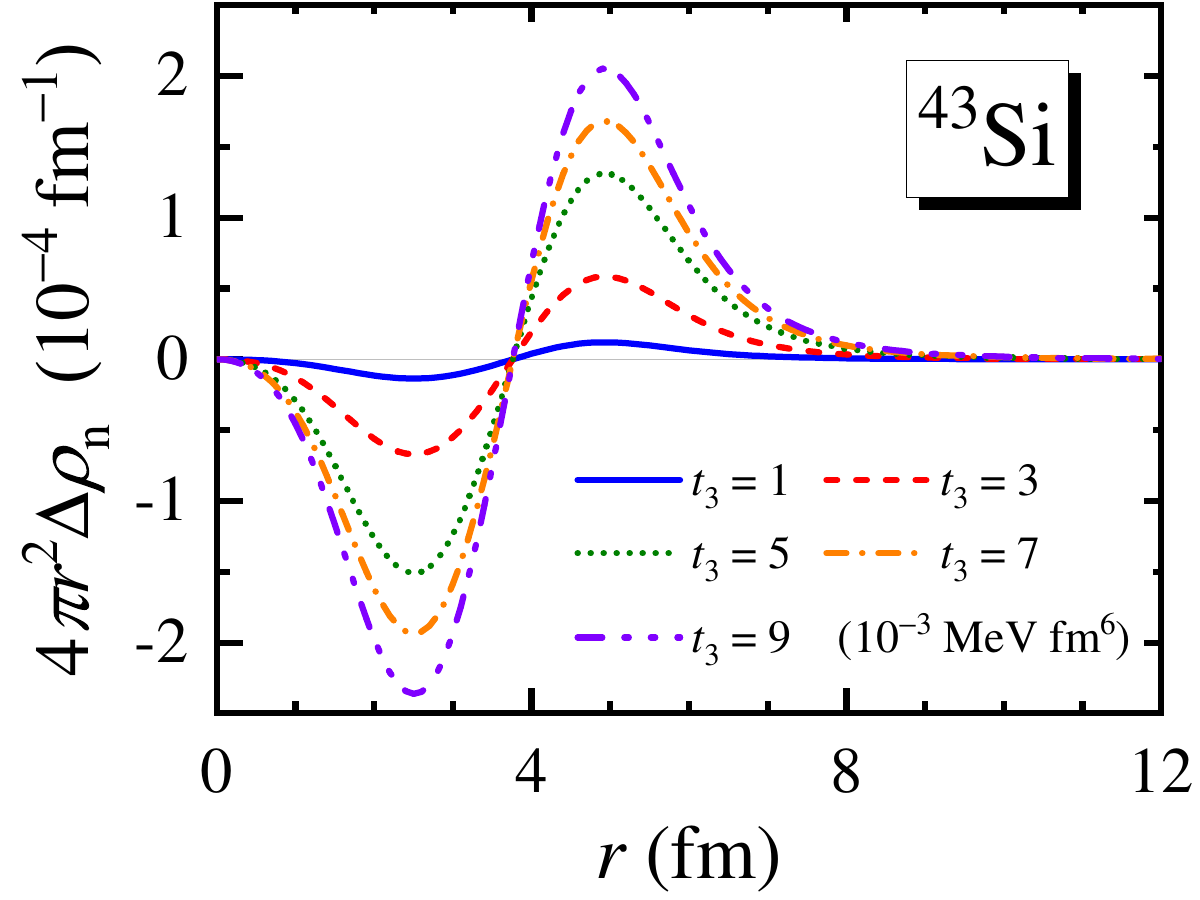}
\caption{Difference of angle average neutron density profiles between the exploratory DRHBc calculations with and without phenomenological three-body force for $^{43}$Si. }
\label{TBF_diff_rho}
\end{figure}

Figure \ref{TBF_diff_rho} shows the influence of three-body force on the neutron density profile for $^{43}$Si. 
As $t_3$ increases, the density at $r<3.8$ fm decreases monotonically, while the density at $r>3.9$ fm increases monotonically. 
This crossing point is very close to the neutron rms radius of $^{43}$Si ($R_{n}=3.898$ fm). 
Therefore, the three-body force pushes neutrons outward from the core, leading to a more diffuse neutron distribution, which is consistent with the increase in neutron radius shown in Fig.~\ref{TBF_diff_E}. 
Thus, the neutron halo is further enhanced by the three-body force.

\section{Framework of the Glauber model}

In our study, with the density of the core nucleus and the wave function of the valence nucleon extracted from the DRHBc theory, reaction observables, $\it{i.e.}$, reaction cross section and momentum distribution can be obtained by the Glauber model \cite{supOgawa1992NPA, supabu2003CPC, supZhang2022JPG, supZhong2022SCP, supAn2024PLB, supWang2024EPJA, supAn2025JPG}. 
Here we outline the theoretical framework for the Glauber model.

The reaction cross section of projectile-target and core-target are defined, respectively, by \cite{supabu2003CPC}: 
\begin{align}
	\sigma_{R}(P+T) & =
	\int \left(1-\left| \Braket{ \varphi_0 \left| e^{{\rm i}\chi _{CT}(\bm{b}_C) + {\rm i}\chi _{NT}
	(\bm{b}_C + \bm{s})} \right| \varphi_0 } \right|^2 \right)\,{\rm d} \bm{b}, \\
	\sigma_{R}(C+T) & =\int \left( 1- \left| e^{{\rm i} \chi _{CT}(\bm{b})} \right|^2 \right)\,{\rm d} \bm{b},
\end{align}
where $\varphi_0$ is the valence-nucleon wave function, $\bm{b}_C = \bm{b} - \bm{s}/A_P$ is the impact parameter between core and target, and $\bm{s}$ is from the coordinate $\bm{r} = (\bm{s},z)$ of the valence nucleon measured from the center-of-mass of the core. 
The core-target phase shift function $\chi_{CT}(\bm{b})$ and nucleon-target phase shift function $\chi_{NT}(\bm{b})$ can be expressed as
\begin{gather}
    {\rm i} \chi_{CT}(\bm{b}) = \int q\rho_C(q) \rho_T(q) f_{NN}(q) J_0(qb) {\rm d} q, \\
    {\rm i} \chi_{NT}(\bm{b}) = \int q\rho_T(q) f_{NN}(q) J_0(qb) {\rm d} q,
\end{gather}
with zero-order Bessel function of first kind $J_0$ and Fourier-transformed densities, $\rho_{C}(q)$ and $\rho_{T}(q)$, where $q$ refers to the one-dimensional transferred momentum from the target to the core. 
In such a way, the core (target) densities $\rho_{C}$ ($\rho_{T}$) can be directly provided by the DRHBc theory. 
The nucleon-nucleon scattering amplitude $f_{NN}(q)$ is parameterized as
\begin{equation}
    f_{NN}(q) = \frac{k_{NN}}{4\pi} \sigma_{NN} ({\rm i} +\alpha_{NN}) \exp \left({\frac{-q^2\beta_{NN}}{2}} \right) ,
\end{equation}
where $k_{NN}$ is the nucleon momentum in the two-nucleon center-of-momentum system, and parameters $\sigma_{NN}$, $\alpha_{NN}$, and $\beta_{NN}$ are derived by taking isospin averages of energy-dependent values $\sigma_{pp}$, $\sigma_{np}$, $\alpha_{pp}$, $\alpha_{np}$, $\beta_{pp}$, and $\beta_{np}$: 
\begin{align}
\sigma_{NN} & = \frac{N_{P}N_{T}\sigma_{nn}+Z_{P}Z_{T}\sigma_{pp}+N_{P}Z_{T}\sigma_{np}+N_{T}Z_{P}\sigma_{np}}{A_{P}A_{T}}, \\
\alpha_{NN} & = \frac{N_{P}N_{T}\alpha_{nn}+Z_{P}Z_{T}\alpha_{pp}+N_{P}Z_{T}\alpha_{np}+N_{T}Z_{P}\alpha_{np}}{A_{P}A_{T}}, \\
\beta_{NN} & = \frac{N_{P}N_{T}\beta_{nn}+Z_{P}Z_{T}\beta_{pp}+N_{P}Z_{T}\beta_{np}+N_{T}Z_{P}\beta_{np}}{A_{P}A_{T}}.
\end{align}
The empirical formulae for $\sigma_{nn}$ and $\sigma_{np}$, are given by~\cite{supCharagi1990PRC}: 
\begin{align}
    \sigma_{nn} = \sigma_{pp} & = 13.73-\frac{15.04}{\beta} + \frac{8.76}{\beta^2}+68.67\beta^{4}, \\
    \sigma_{np} & = -70.67-\frac{18.18}{\beta}+\frac{25.26}{\beta^2}+113.85\beta,
\end{align}
where $\beta=v/c$, $v$ refers to the speed of the projectile nucleus, and $c$ is the speed of light. 
The values of $\alpha_{pp}$, $\alpha_{np}$, $\beta_{pp}$, and $\beta_{np}$ used in our calculations are taken from Ref.~\cite{supAbu2008PRC} and listed in Table.~\ref{tab:parameters}.

\begin{table}[htbp]
    \centering
    \caption{Parameters $\alpha_{pp}$, $\alpha_{np}$, $\beta_{pp}$ and $\beta_{np}$ at the energy of 240 MeV~\cite{supAbu2008PRC}.}
    \begin{tabular}{ccccc}
        \hline
        Energy (MeV) & $\alpha_{pp}$ & $\alpha_{np}$ & $\beta_{pp}$ (fm$^{2}$) & $\beta_{np}$ (fm$^{2}$) \\
        \hline
        240 & 0.944 & 0.541 & 0.086 & 0.106 \\
        \hline
    \end{tabular}
    \label{tab:parameters}
\end{table}

Another critical reaction observable to judge halo nuclei is the longitudinal momentum distribution of the valence nucleon and the core nucleus after one-neutron removal reaction, which is defined as~\cite{supabu2003CPC}
\begin{align}
    \label{momentum distribution}
    \begin{aligned}
    \frac{{\rm d}\sigma_{-N}^{inel} }{{\rm d}\bm{p}_{\parallel} } = & \frac{1}{2\pi\hbar} \int {\rm d}\bm{b}_N\left[1-e^{-2{\rm Im}\chi_{NT}(\bm{b}_N)}\right] \int {\rm d}\bm{s}e^{-2 {\rm Im} \chi_{CT}(\bm{b}_N - \bm{s})} \\ 
    & \times \int {\rm d}z\int {\rm d}{z'}e^{\frac{\rm i}{\hbar}\bm{p}_{\parallel}(z-{z'})}u_{nlj}^*(r')u_{nlj}(r)\frac{1}{4\pi}P_l({\hat{\bm{r}}}\cdot{\hat{\bm{r'}}}),
    \end{aligned}
\end{align}
in which $\bm{r}=(\bm{s},z)$ and $\bm{r'}=(\bm{s},z')$ are the coordinates of the valence nucleon measured from the center-of-mass of the core, $\bm{b}_N$ stands for the impact parameter of the valence nucleon with respect to the target, $u_{nlj}$ is the radial wavefunction of the valence-nucleon, and $P_l$ denotes $l$-order Legendre polynomial.

The density of target $^{12}$C takes the available form of two sets of Gaussian functions obtained by fitting results of the harmonic-oscillator shell model, which can reproduce the measured interaction cross section~\cite{supOgawa1992NPA}. 
More details on the Glauber model can be found in Refs.~\cite{supZhang2022JPG, supZhong2022SCP, supAn2024PLB, supWang2024EPJA, supAn2025JPG} and the references therein.


\begin{figure}[htbp]
\centering
\includegraphics[width=0.8\linewidth]{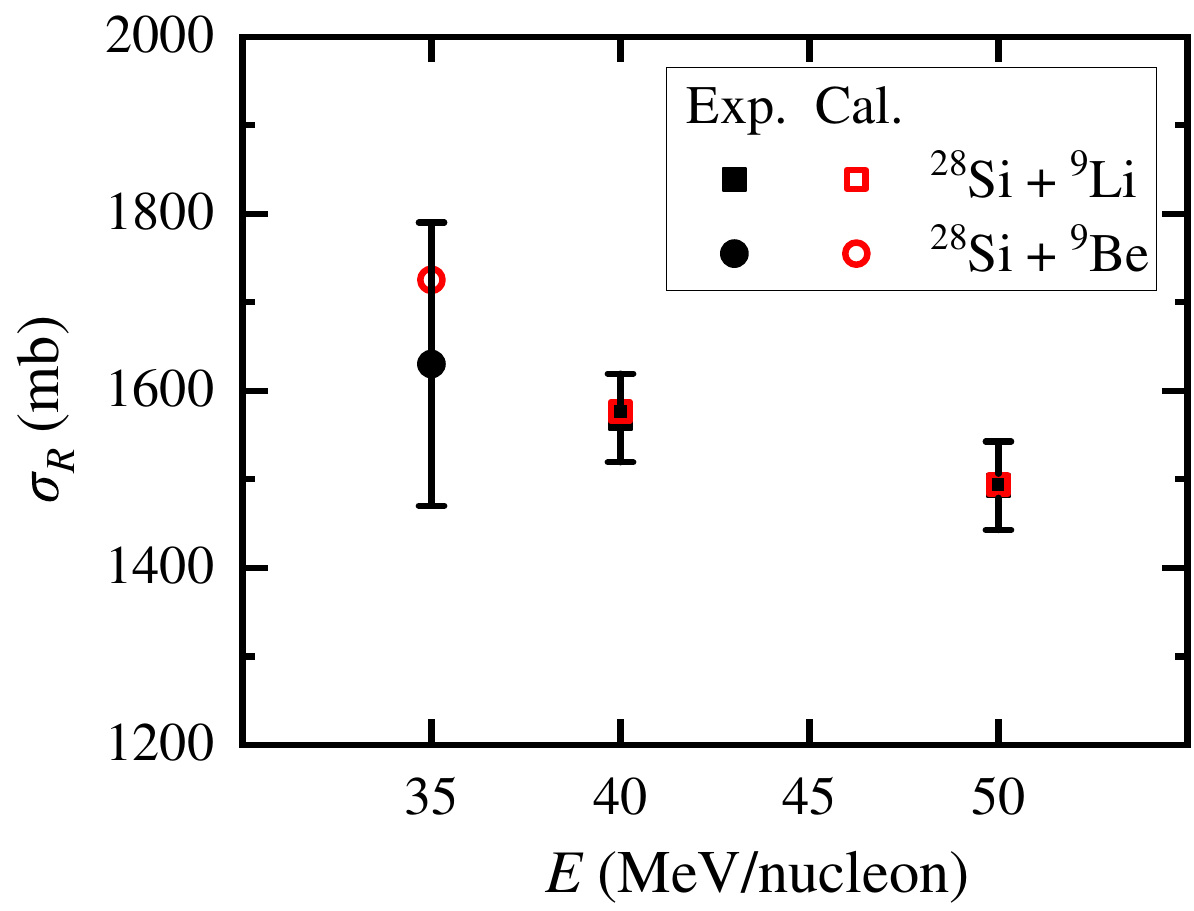}
\caption{Reaction cross sections of $^{28}$Si bombarding $^{9}$Li and $^{9}$Be targets evaluated using the Glauber model with inputs from the DRHBc model, in comparison with the available data from Refs. \cite{supHue2017JPCS,supPenionzhkevich2018KnE}. }
\label{glauber_Si28}
\end{figure}

To validate the reliability of the present approach, we calculated the reaction cross sections of $^{28}$Si bombarding $^{9}$Li and $^{9}$Be targets based on the Glauber model with inputs from the DRHBc model.
The results are compared with available data from Refs. \cite{supHue2017JPCS,supPenionzhkevich2018KnE}, as shown in Fig.~\ref{glauber_Si28}. 
The calculated results are in good agreement with all available data within the experimental uncertainties. 
No adjustable free parameter is introduced in our calculations. 


\end{document}